\documentclass[conference]{IEEEtran}
\IEEEoverridecommandlockouts
\usepackage[letterpaper, left=.625in, right=.625in, bottom=1in, top=0.7in]{geometry}
\usepackage{cite}
\usepackage{amsmath,amssymb,amsfonts}
\usepackage[bookmarks,colorlinks]{hyperref} 
\hypersetup{colorlinks,citecolor= red,filecolor= blue,linkcolor= blue,urlcolor=blue}
\usepackage{graphicx}
\usepackage{subcaption}
\usepackage{float}
\usepackage{textcomp}
\usepackage{xcolor}
\usepackage{algorithm}
\usepackage{algpseudocode}
\usepackage{balance}
\usepackage[font=small,labelfont=bf]{caption}

\def\BibTeX{{\rm B\kern-.05em{\sc i\kern-.025em b}\kern-.08em
		T\kern-.1667em\lower.7ex\hbox{E}\kern-.125emX}}

\newcommand{\bd}{\begin{description}}
\newcommand{\ed}{\end{description}}
\newcommand{\be}{\begin{enumerate}}
\newcommand{\ee}{\end{enumerate}}
\newcommand{\bi}{\begin{itemize}}
\newcommand{\ei}{\end{itemize}}
\newcommand{\bl}{\begin{list}}
\newcommand{\el}{\end{list}}
\newcommand{\bt}{\begin{tabbing}}
\newcommand{\et}{\end{tabbing}}

\definecolor{BLUE}{rgb}{0,0,1}

\usepackage{acronym}  
\acrodef{siso}[SISO]{single-input single-output}
\acrodef{csi}[CSI]{channel state information}
\acrodef{bem}[BEM]{basis expansion model}
\acrodef{ls}[LS]{least squares}
\acrodef{mmse}[MMSE]{minimum mean square error}
\acrodef{pa}[PA]{pilot-aided}
\acrodef{dd}[DD]{decision-directed}
\acrodef{ce}[CE]{channel estimation}
\acrodef{mse}[MSE]{mean square error}
\acrodef{dnn}[DNN]{deep neural network}
\acrodef{cnn}[CNN]{convolutional neural network}
\acrodef{mse}[MSE]{mean-squared error}
\acrodef{dl}[DL]{deep learning}
\acrodef{ci}[CI]{channel state information}
\acrodef{mmse}[MMSE]{minimum mean square error}
\acrodef{awgn}[AWGN]{additive white Gaussian noise}
\acrodef{map}[MAP]{maximum a posteriori probability}
\acrodef{ber}[BER]{bit error rate}
\acrodef{aer}[AER]{activity error rate}
\acrodef{kf}[KF]{Kalman filter}
\acrodef{snr}[SNR]{signal-to-noise ratio}
\acrodef{lse}[LSE]{least-squared error}
\acrodef{flops}[Flops]{floating-point operations}
\acrodef{ml}[ML]{Maximum likelihood}
\acrodef{mdp}[MDP]{Markov decision process}
\acrodef{rl}[RL]{reinforcement learning}
\acrodef{mud}[MUD]{multiuser detection}
\acrodef{cdma}[CDMA]{code-division multiple access}
\acrodef{cs}[CS]{compressive sensing}
\acrodef{iot}[IoT]{Internet of Things}
\acrodef{m2m}[M2M]{machine-to-machine}
\acrodef{mmtc}[mMTC]{massive machine-type communications}
\acrodef{5g}[5G]{fifth-generation}
\acrodef{noma}[NOMA]{non-orthogonal multiple access}
\acrodef{bs}[BS]{base station}
\acrodef{pdf}[PDF]{probability density function}
\acrodef{tpr}[TPR]{true possitive rate}
\acrodef{fpr}[FPR]{false possitive rate}
\acrodef{roc}[ROC]{receiver operating characteristic}
\acrodef{omp}[OMP]{orthogonal matching pursuit}
\acrodef{amp}[AMP]{approximate message passing}
\acrodef{mtc}[MTC]{machine-type communication}
\acrodef{fl}[FL]{federated learning}
\acrodef{ai}[AI]{Artificial Intelligence}
\acrodef{air}[AirComp]{over-the-air computation}
\acrodef{fedavg}[FedAvg]{Federated Averaging}
\acrodef{sgd}[SGD]{stochastic gradient descent}
\acrodef{mlp}[MLP]{multilayer perceptron}
\acrodef{ad}[AD]{activity detection}

\usepackage[yyyymmdd,hhmmss]{datetime} 
\newdateformat{monthyeardate}{\monthname[\THEMONTH] \THEDAY, \THEYEAR} 

\begin{document}
	\title{Activity Detection for Grant-Free NOMA in Massive IoT Networks
	}
	\author{\IEEEauthorblockN{
		Mehrtash Mehrabi, \textit{Student Member, IEEE},~Mostafa~Mohammadkarimi, \textit{Member, IEEE},\\and Masoud~Ardakani,~\textit{Senior~Member,~IEEE}}
	\IEEEauthorblockA{Department of Electrical and Computer Engineering, University of Alberta, Edmonton, AB, T6G 1H9, Canada}
	Email: $\lbrace \text{mehrtash, mostafa.mohammadkarimi, ardakani}\rbrace$@ualberta.ca}

	\maketitle
	
\begin{abstract}
    Recently, grant-free transmission paradigm has been introduced for massive \ac{iot} networks to save both time and bandwidth and transmit the message with low latency. 
    In order to accurately decode the message of each device at the \ac{bs}, first, the active devices at each transmission frame must be identified.
    In this work, first we investigate the problem of activity detection as a threshold comparing problem. 
    We show the convexity of the activity detection method through analyzing its probability of error which makes it possible to find the optimal threshold for minimizing the activity detection error.
    Consequently, to achieve an optimum solution, we propose a \ac{dl}-based method called \ac{cnn}-\ac{ad}.
    In order to make it more practical, we consider unknown and time-varying activity rate for the \ac{iot} devices.
	Our simulations verify that our proposed  CNN-AD method can achieve higher performance compared to the existing non-Bayesian greedy-based methods.
	This is while existing methods need to know the activity rate of IoT devices, while our method works for unknown and even time-varying activity rates.
\end{abstract}


	\IEEEpeerreviewmaketitle
	\section{Introduction}
	
	\acresetall
	
	\IEEEPARstart{W}{ireless} technology recent advances provide massive connectivity for machines and objects resulting in the \ac{iot} \cite{IoTint}.
	The demand for the \ac{iot} is expected to grow drastically in the near future with numerous applications in health care systems, education, businesses and governmental services \cite{IoTInd, iot1, MassiveMachineType}.
	
	As the demand for connectivity in \ac{iot} systems is growing  rapidly, it is crucial to improve the spectrum efficiency \cite{iotcapacity}.
	Hence, the \ac{noma} has been introduced \cite{noma}.
	To address the main challenges of \ac{iot}, including access collisions and massive connectivity, \ac{noma} allows devices to access the channel non-orthogonally by either power-domain \cite{PowerNoma} or code-domain \cite{SCMANOMA} multiplexing.
	Meanwhile, this massive connectivity is highly affected by the conventional grant-based \ac{noma} transmission scheme, where the exchange of control signaling between the \ac{bs} and \ac{iot} devices is needed for channel access.
	The excessive signaling overhead causes spectral deficiency and large transmission latency.
	Grant-free \ac{noma} has been introduced to make a flexible transmission mechanism for the devices and save time and bandwidth by removing the need for the exchange of control signaling between the \ac{bs} and devices.
	Hence, devices can transmit data randomly at any time slot without any request-grant procedure.	
	
	In many \ac{iot} applications, a few devices become active for a short period of time to communicate with the \ac{bs} while others are inactive \cite{IoTMag}.
	In \ac{iot} networks with a large number of nodes each with a small probability of activity, \ac{mud} methods heavily rely on \ac{ad} prior to detection and decoding \cite{MassiveMachineType,verdu1998multiuser, zhang2018block, GiaSparseActivityMUD, CDMAMUD}.
	For uplink transmission in \ac{iot} systems with grant-free \ac{noma} transmission scheme, where the performance of \ac{mud} can be severely affected by the multi-access interference, the reliable detection of both activity and transmitted signal is very challenging and can be computationally expensive \cite{verdu1998multiuser,GiaSparseActivityMUD}.
	
	There have been many studies in the literature suggesting \ac{cs} methods for joint activity and data detection \cite{GiaSparseActivityMUD,CDMAMUD,SparseActivityDetection,SparseCDMA, wang2020compressive}.
	Although \ac{cs} methods can achieve a reliable \ac{mud}, they only work in networks with sporadic traffic pattern, and are expensive in terms of computational complexity \cite{GiaSparseActivityMUD}.
	Recently, \ac{dl} models have observed a lot of interests in communication systems and more specifically in signal detection \cite{deeplearning, DLSphere, NOMAActivity}.
	A study in \cite{NOMAActivity} suggests to use \ac{dl} for activity and data detection, however they consider a deterministic traffic pattern for the activity which is not valid in all environments.

	In this work, we first formulate the problem of IoT activity detection as a threshold comparing problem. We then analyze the probability of error of this activity detection method. Observing that this probability of error is a convex function of the decision threshold, we raise the question of finding the optimal threshold for minimizing the activity detection error. To achieve this goal, we propose a \ac{cnn}-based AD algorithm for grant-fee code-domain uplink NOMA. Unlike existing CS-based AD algorithms, our solution  does not need to know the exact number of active devices or even the activity rate of IoT devices. In fact, in our system model we assume a time-varying and unknown activity rate and a heterogeneous network. Simulation results verify the success the proposed algorithm.

	The rest of this paper is organized as follows.
	We present the system model in Section \ref{Sec:system}.
	In Section \ref{Sec:DetectorAlg} we formulate the device \ac{ad} problem and derive its probability of error. 
	Section \ref{Sec:method} introduces our \ac{cnn}-based solution for device \ac{ad}.
	The simulation results are presented in Section \ref{Sec:simulations}.
	Finally, the paper is concluded in Section \ref{Sec:Conclusions}.
	
	\subsection{Notations}
	Throughout this paper, $(\cdot)^*$ represents the complex conjugate. 
	Matrix transpose and Hermitian operators are shown by $(\cdot)^T$ and $(\cdot)^H$, respectively.
	The operator $\text{diag}(\mathbf{b})$ returns a square diagonal matrix with the elements of vector $\mathbf{b}$ on the main diagonal.
	Furthermore, $\mathbb{E}[\cdot]$ is the statistical expectation, $\hat{\mathbf{a}}$ denotes an estimated value for $\mathbf{a}$, and size of set $\mathcal{S}$ is shown by $|\mathcal{S}|$.
	The constellation and $m$-dimensional complex spaces are denoted by $\mathbb{D}$ and $\mathbb{C}^m$, respectively.
	Finally, the circularly symmetric complex Gaussian distribution with mean vector $\mathbf{\mu}$ and covariance matrix $\mathbf{\Sigma}$ is denoted by $\mathcal{CN}(\mathbf{\mu},\mathbf{\Sigma})$.
\section{System Model}
\label{Sec:system}
\begin{figure}[t]
    \vspace{-.2cm}
    \centering
    \includegraphics[width=0.5\textwidth]{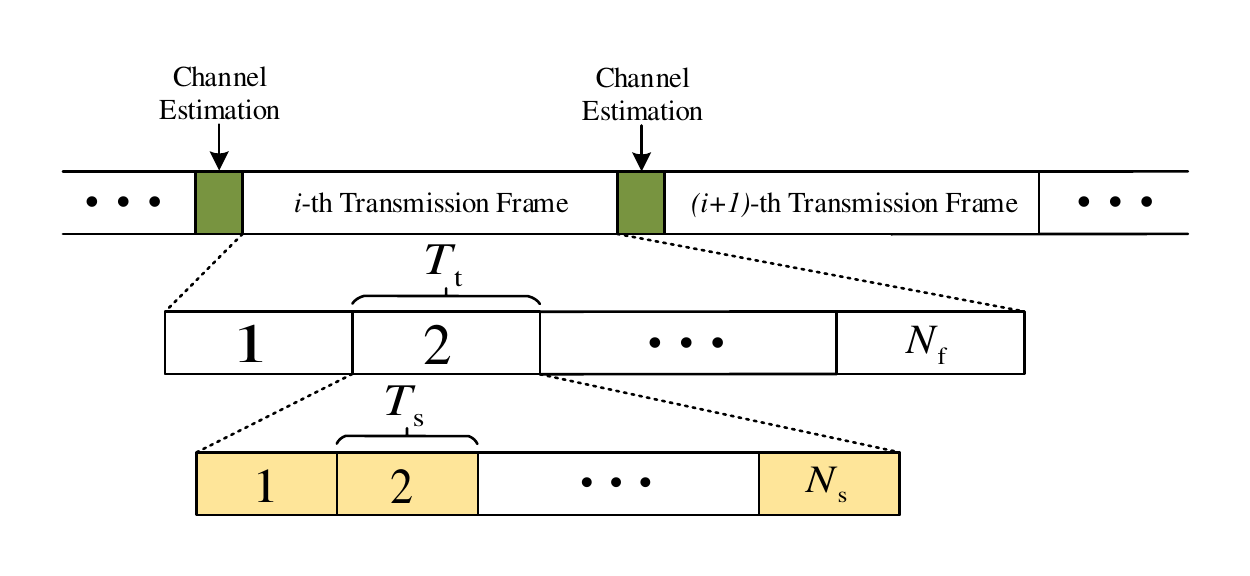}
    \vspace{-.6cm}
    \caption{CDMA slotted ALOHA transmission frame}
    \vspace{-.5cm}
    \label{fig:transmissinoFrame}
\end{figure}

    We consider a \ac{cdma} uplink  transmission, where $K$ \ac{iot} devices communicate with a single \ac{iot} \ac{bs} equipped with $M$ receive antennas.
    This commonly used model \cite{iot1,noma,NOMAActivity}, also considers a frame structure for uplink transmission composed of a channel estimation phase followed by CDMA slotted ALOHA data transmissions as shown in Fig.~\ref{fig:transmissinoFrame}.  
    In each frame, let $N_{\rm f}$ short packets of length $T_{\rm t}=N_{\rm s}T_{\rm s}$, where $N_{\rm s}$ is the number of symbols per \ac{iot} packet and $T_{\rm s}$ is the symbol duration.
    It is assumed that the channel is fixed during each frame, but it varies from one frame to another.
    The \ac{csi} is acquired at the \ac{bs} during the channel estimation phase.
    As it is common in \ac{mmtc}, we assume that
    the \ac{iot} devices are only active on occasion and transmit short data packets during each frame. 
    The activity rate of the \ac{iot} devices is denoted by
    $P_{\rm a}\in[0,P_{\rm max}]$, which is assumed to be unknown and time-varying from one packet transmission to another.
Let $b_k\in\cal A$ be the transmitted symbol of the $k$-th device and chosen from a finite alphabet $\cal A$, when the $k$-th device is active; otherwise, $b_k=0$.
Consequently, $b_k$ can take values from an augmented alphabet $\bar{{\cal A}}={\cal A}\cup\{0\}$.
We also denote the set of all devices and the set of active devices by ${\cal S}_{\rm t}=\{1,2,\dots,K\}$ and
${\cal S}_{\rm a}$, respectively, where ${\cal S}_{\rm a} \subset {\cal S}_{\rm t}$.\footnote{For the simplicity of notation, we remove the index of frame and packet.}

A unique spreading code is dedicated to each
\ac{iot} device which is simultaneously used for the spreading
purpose and device identification. 	This removes the need for control signaling associated with IoT
device identification.
Control signals are inefficient for short packet \ac{mmtc}.
The spreading sequence for the $k$-th \ac{iot} device is denoted by $\mathbf{c}_k=[c_{1,k} ~~ c_{2,k} ~~\cdots ~~c_{N_c,k}]^T$ where $c_{i,k}\in\{-1,+1\}$ and $N_{\rm c}$ is the spreading factor. 
To support a large number of devices, non-orthogonal spreading sequences are employed; resulting in \ac{noma} transmission.

For a single frame, the complex channel coefficient between the $k$-th \ac{iot} device and the $m$-th receive antenna at the \ac{bs} is denoted as $g_{m,k}$.
The active \ac{iot} device $k$, $k\in{\cal S}_{\rm a}$ transmits $N_{\rm s}$ symbols denoted by $\mathbf{b}_k = [b_{k,1},\cdots,b_{k,N_{\rm s}}]^T$ during a packet.
	The received baseband signal over Rayleigh flat fading channel in a single slot of the slotted ALOHA frame at the $m$-th receive antenna of the \ac{bs} is expressed as
	\begin{equation}
		\mathbf{Y}_m=\sum_{k=1}^{K}g_{m,k}\mathbf{c}_k\mathbf{b}_k^T+\mathbf{W}_{m},
	\end{equation}
	where $\mathbf{W}_{m}\in \mathbb{C}^{N_{\rm c}\times N_{\rm s}}$ with $w_{i,j}\sim {\cal CN}(0,\sigma_{\rm w}^2)$ and $E[w_{i,j}w_{u,v}^*]=\sigma_{\rm w}^2 \delta[i-u]\delta[j-v]$ is the \ac{awgn} matrix at the $m$-th receive antenna.
	The equivalent channel matrix between all IoT devices and the $m$-th receive antenna can be expressed as $\mathbf{\Phi}_m=[g_{m,1}\textbf{c}_1, \cdots, g_{m,K}\textbf{c}_{K}]\in\mathbb{C}^{N_{\rm c}\times K}$.
	Thus, the received packet at the $m$-th ($m=1,2,\cdots,M$) receive antenna is given by 
	\begin{equation}
		\mathbf{Y}_m = \mathbf{\Phi}_m\textbf{B} + \mathbf{W}_{m},
	\end{equation}
	where $\textbf{B} = [\mathbf{b}_1, \cdots, \mathbf{b}_{K}]^T\in\mathbb{D}^{K\times N_{\rm s}}$. 
	
	Let the total set of all \ac{iot} devices be decomposed into a finite number of disjoint groups $\mathcal{G}_1,\mathcal{G}_2,\cdots,\mathcal{G}_S$.
    Within group $\mathcal{G}_j$, the
    power of every \ac{iot} device is given by $P_j$.
    The powers of the devices are equal within each group, but differ from group to group.
    The fraction of devices in group $\mathcal{G}_j$ is therefore $|\mathcal{G}_j|/K$.
    It is assumed that $P_j$ is known at the \ac{bs}. 
    This configuration captures heterogeneous \ac{iot} networks, where groups of \ac{iot} devices capture different phenomenon in a given geographical area. 
    A single group of \ac{iot} devices with equal power transmission, resulting in a homogeneous network, is also studied in this paper.

\section{Problem Formulation} \label{Sec:DetectorAlg}
In this section, we present the problem of \ac{iot} device \ac{ad} in the cases of known \ac{csi} at the receiver and in the presence of sparse or non-sparse transmission. 
In order to detect the active devices, it is assumed that the \ac{bs} is equipped with a match filter and the precoding matrix and \ac{csi} $\mathbf{\Phi}_m$ is available.
Before \ac{ad}, the observation matrix at the $m$-th receive antenna $\mathbf{y}_m$ is passed through the decorrelator to obtain
\begin{equation}
	\mathbf{\overline{Y}}_{m} = \mathbf{\Phi}_m^H\mathbf{Y}_m \in \mathbb{C}^{K\times N_{\rm s}}.
\end{equation}
In the following, we investigate the details of the \ac{ad} problem based on the Gaussian detection to show how a threshold can be computed to distinguish active \ac{iot} devices from inactive ones.
The output of the decorrelator receiver for the $m$-th receive antenna is expressed as
	\begin{align}\nonumber
	&	\mathbf{\overline{Y}}_{m} = \mathbf{\Phi}_m^H\mathbf{\Phi}_m\mathbf{B}+\mathbf{\Phi}_m^H\mathbf{W}_m, \\
		&= \begin{bmatrix}
			\sum_{k=1}^{K}g_{m,1}^*g_{m,k}\mathbf{c}_1^T\mathbf{c}_k\mathbf{b}_k^T+g_{m,1}^*\mathbf{c}_1^T\mathbf{W}_m \\
			\sum_{k=1}^{K}g_{m,2}^*g_{m,k}\mathbf{c}_2^T\mathbf{c}_k\mathbf{b}_k^T + g_{m,2}^*\mathbf{c}_2^T\mathbf{W}_m \\
			\vdots \\
			\sum_{k=1}^{K}g_{m,K}^*g_{m,k}\mathbf{c}_{K}^T\mathbf{c}_k\mathbf{b}_k^T + g_{m,K}^*\mathbf{c}_{K}^T\mathbf{W}_m
		\end{bmatrix}.
	\end{align}
	Consequently, the received signal from the $k$-th user at the $m$-th receive antenna is
	\begin{equation}
		\mathbf{r}_{k}^{m} = ||g_{m,k}\mathbf{c}_k||_2^2 \mathbf{b}_k^T + \sum_{i=1 (i\ne k)}^{K} g_{m,k}^*g_{m,i} \mathbf{c}_{k}^T\mathbf{c}_i\mathbf{b}_i^T +g_{m,k}^*\mathbf{c}_k^T\mathbf{W}_m,
	\end{equation}
	where the second and third terms are multi user interference and additive noise, respectively.
	Since an \ac{iot} device is either active or inactive for the entire packet transmission, we determine the activity status of a device based on each received symbol and then use the results in \cite{hardcombine} for spectrum sensing and combine the obtained results from all $N_s$ symbols.
	The device \ac{ad} in the case of single symbol transmission is studied in \cite{GiaSparseActivityMUD}, and we follow that to determine the status of each device based on each received symbol and then combine the results.
	The $j$-th received symbol from device $k$ at receive antenna $m$, denoted as  $r_{k,j}^m$, is
	\begin{align}\nonumber
		r_{k,j}^m =& ||g_{m,k}\mathbf{c}_k||_2^2 b_{k,j} + \\ &\sum_{i=1 (i\ne k)}^{K} g_{m,k}^*g_{m,i}\mathbf{c}_{k}^T\mathbf{c}_ib_{i,j} + g_{m,k}^*\mathbf{c}_k^T\mathbf{w}_j,
	\end{align}
	where the first term is the main signal, the second term is multi user interference from other devices, and the third term is the additive noise.
	For the sake of simplicity we assume that BPSK modulation is used, i.e., the transmitted symbols are drawn from ${\cal A}=\{-1,+1\}$ and $p(b_{k,j} = +1)=p(b_{k,j} = -1)=1/2$.
	The multi user interference plus noise in $r_{k,j}^m$ has variance
	\begin{align}\nonumber
		\sigma^2_{k,j} = & \text{~var}\Big{\{}\sum_{i=1 (i\ne k)}^{K} g_{m,k}^*g_{m,i}\mathbf{c}_{k}^T\mathbf{c}_ib_{i,j} + g_{m,k}^*\mathbf{c}_k^T\mathbf{w}_j\Big{\}} \\
		= & \sum_{i=1 (i\ne k)}^{K}|g_{m,k}^*g_{m,i}\mathbf{c}_{k}^T\mathbf{c}_i|^2 P_a + ||g_{m,k}^*\mathbf{c}_k^T||_2^2.
	\end{align}
	

	Now we can approximate $r_{k,j}^m$ by a Gaussian distribution as ${\cal N}(||g_{m,k}\mathbf{c}_k||_2^2 ,\sigma^2_{k,j})$ \cite{hardcombine}.
	In order to identify the activity of device $k$, our goal is to propose an algorithm to define threshold $\tau$ and set device $k$ as active if $|r_{k,j}^m|>\tau$.
	Then the probability of error, $P_e$, is computed as
	\begin{align}\label{eq:error}\nonumber
		P_e^{k,j} =& P_a p(|r_{k,j}^{m}|<\tau|b_{k,j} \ne 0) \\&+ 2(1-P_a)p(|r_{k,j}^{m}|>\tau|b_{k,j} = 0),
	\end{align}
	where we have $p(r_{k,j}^{m}|b_{k,j} \ne 0)\sim{\cal N}(||g_{m,k}\mathbf{c}_k||_2^2,\sigma^2_{k,j})$ and $p(r_{k,j}^{m}|b_{k,j} = 0)\sim{\cal N}(0,\sigma^2_{k,j})$.
	We can rewrite \eqref{eq:error} as
	\begin{align}\label{eq:errorQ}
		P_e^{k,j} = 2(1-P_a) Q(\frac{\tau}{\sigma_{k,j}}) + P_a Q(\frac{||g_{m,k}\mathbf{c}_k||_2^2-\tau}{\sigma_{k,j}}),
	\end{align}
	where $Q(x)=(1/\sqrt{2\pi})\int_{x}^{\infty}\exp(-t^2/2)dt$ denotes the Gaussian tail function.
	The probability of error in \eqref{eq:errorQ} is a convex function of $\tau$ and hence, a fine tuned neural network is capable of solving this problem and detect the active devices by finding the optimum $\tau$.
	In the following section, we define our \ac{dl}-based algorithm to find the optimum $\tau$ and minimize the probability of error.

	\begin{figure*}
    \centering
    \includegraphics[width=.95\textwidth]{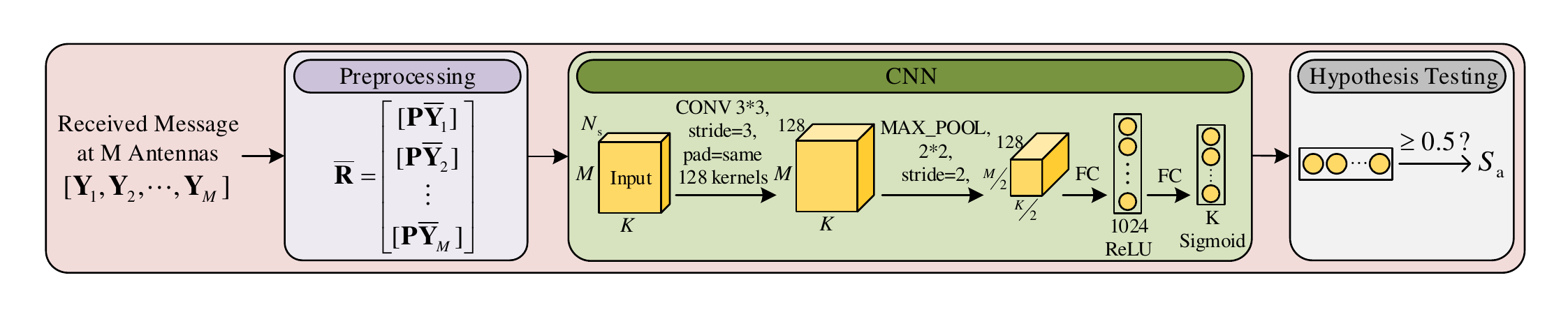}
    \vspace{-.4cm}
    \caption{Model structure of the proposed CNN-AD algorithm}
    \vspace{-.4cm}
    \label{fig:modelStructure}
    \end{figure*}
\section{DL-Based AD}\label{Sec:method}
	Device \ac{ad} is the first step toward effective \ac{mud} in a grant-free uplink multiple access.  
	The recent studies on \ac{ad} suggest to use \ac{cs} methods to identify the set of active devices \cite{SparseActivityDetection,SparseCDMA}. 
	However, these methods fail in the practical scenarios, where the activity rate is time-varying and/or unknown. 
	Moreover, these methods are mainly effective for low device activity rate scenarios, i.e., when sparsity level is high \cite{SparseActivityDetection}.
    In this section, we propose our \ac{ad} algorithms called CNN-AD by employing a \ac{cnn} for heterogeneous IoT networks. 
    By employing a suitably designed \ac{cnn}, the underlying pattern in device activity can be easily learnt.

\subsection{\ac{cnn}-\ac{ad} Algorithm}
	Fig. 2 illustrates the structure of the proposed CNN-AD algorithm. As seen, it is composed of there blocks: 1) preprocessing, 2) CNN processing, and 3) hypothesis testing. 
	
	In the preprocessing step after sequence matched filtering, 
	we first sort the observation matrix from all $M$ receive antennas in a 3D Tensor  as 
	\begin{align}\label{eqtttu}
{\mathbf{\overline{R}}} = \left[ \begin{array}{l}
\left[ {{\bf{P}}{{{\bf{\bar Y}}}_1}} \right]\\
\left[ {{\bf{P}}{{{\bf{\bar Y}}}_2}} \right]\\
\,\,\,\,\, \vdots \\
\left[ {{\bf{P}}{{{\bf{\bar Y}}}_M}} \right]
\end{array} \right]
	\end{align}
	where $\mathbf{P\overline{Y}}_m\in \mathbb{C}^{K\times N_{\rm s}}$,
 $\mathbf{\overline{Y}}_{m} = \mathbf{\Phi}_m^H\mathbf{Y}_m \in \mathbb{C}^{K\times N_{\rm s}}$ for $m=1,2,\cdots, M$, and  
	$\textbf{P} \triangleq  {\rm diag}(p_1,\cdots,p_K)$, $p_k\in\{1/P_1,\cdots,1/P_S\}$ for $k=1,2,\cdots,K$.  
	
	
	In the CNN processing block, the 3D Tensor $\mathbf{\overline{R}}$ is used as the input of a suitable designed \ac{cnn}. The \ac{cnn} models benefit from the convolutional layers performing convolution operations between matrices instead of multiplication. Thus, 
	it leads to dimension reduction for feature extraction and provides a new input to the next network layers which includes only the useful features of the original high-dimensional input.
    The IoT device \ac{ad} can be formulated as a binary classification or regression problem. Formulating device \ac{ad} as a classification problem is straightforward, but it requires the accurate design of the \ac{cnn}'s structure and parameters. 
	
	In the hypothesis testing block, the $K$ 
	outputs of the \ac{cnn}'s  Sigmoid layer is compared with a predefined threshold to determine the activity status of the \ac{iot} devices in the network. If the $k$-th node of the Sigmoid layer exceeds the threshold, the $k$-th \ac{iot} device is identified as active. 

    \subsection{Training Phase}    
	In order to train the designed \ac{cnn}, we define the activity vector $\mathbf{a}$ as
	\begin{equation}
		\mathbf{a} = [a_1 ~~ a_2 ~~ \cdots ~~ a_{K}]^T,
	\end{equation}
	where $a_k$ is 1 when the $k$-th \ac{iot} device is active and 0 otherwise.
    We train our model with $N$ independent training samples ($\mathbf{\overline{R}}^{(j)}$,$\mathbf{a}^{(j)}$), where $j=1,2,\cdots,N$ and $\mathbf{a}^{(j)}$ and $\mathbf{\overline{R}}^{(j)}$ are the activity vector and observation matrix of the $j$-th training sample, respectively.
	Our objective is to train the designed \ac{cnn} to generate the desired output vector $\mathbf{a}^{(j)}$ for input matrix $\mathbf{\overline{R}}^{(j)}$.
	The model tries to learns non-linear transformation $\Psi$ such that
	\begin{equation}\label{eq:cnn}
		\mathbf{\hat{a}}^{(j)} = \Psi(\mathbf{\overline{R}}^{(j)};\bf{\Theta}),
	\end{equation}
	where $\bf{\Theta}$ is the set of parameters learned during the training by minimizing the loss function.
	The output of model, i.e. $\mathbf{\hat{a}}$ determines the activity probabilities of the \ac{iot} devices.
	Here since there are two classes (active or inactive) for each \ac{iot} device, the loss function is chosen as the binary cross-entropy.
	For each training sample $j$, the binary cross-entropy loss function compares the probability that the \ac{iot} devices are active ($\mathbf{\hat{a}}^{(j)}$) with the true activity vector $\mathbf{a}^{(j)}$ as 
	\begin{equation}\label{eq:lossFunction}
	\text{Loss}{(\bf{\Theta})}=\frac{1}{N}\sum_{j=1}^{N}
	-\Big(\mathbf{a}^{(j)}\log(\mathbf{\hat{a}}^{(j)}) + ({\bf{1}}-\mathbf{a}^{(j)})\log({\bf 1}-\mathbf{\hat{a}}^{(j)})\Big),	
	\end{equation}
	where $\log(\cdot)$ performs an element-wise $\log$ operation on $\mathbf{\hat{a}}^{(j)}$, and the vector multiplication is also element-wise.

\section{Experiments}
\label{Sec:simulations}
In this section, we evaluate the performance of the proposed CNN-AD algorithm through various simulation experiments and compare it with some of the existing methods.

\subsection{Simulation Setup}
We consider an \ac{iot} network with $K$ devices where $K>N_{\rm c}$ and pseudo-random codes are used as the spreading sequences for \ac{iot} devices.
The probability of activity $P_{\rm a}$ is considered to be unknown and time-varying from one packet to another in the range of $P_{\rm a}\in[0,P_{\rm max}]$, where $P_{\rm max}=0.1$.
The BPSK modulation is used for uplink transmission. Without loss of generality, the channel coefficient between \ac{iot} devices and the \ac{bs} is modeled as independent zero-mean complex Gaussian random variables with variance $\sigma_{k,m}^2=1, k\in {\cal S}_{\rm t}$ and $m\in\{1,\cdots,M\}$.
The additive white noise is modeled as zero-mean complex Gaussian random variables with variance $\sigma_{\rm w}^2$, and the \ac{snr} in dB is defined as $\gamma \triangleq  10\log(\sigma_{\rm{s}}^2/\sigma_{\rm{w}}^2)$, where $\sigma_{\rm{s}}^2=P_{\rm a}P_{\rm t}$ is the average transmit power with $P_{\rm t}=\sum_{k=1}^Kp_k$ as the total transmission power.
Unless otherwise mentioned, we consider spreading sequences with spreading factor $N_{\rm c}=32$.

In order to train CNN-AD, we generate $10^5$ independent samples and use 80\% for training and the rest for validation and test.
Adam optimizer \cite{adam} with learning rate of $10^{-3}$ is used to minimize cross-entropy loss function in \eqref{eq:lossFunction}.



\begin{figure}[t]
	\centering
	\includegraphics[width=0.45\textwidth]{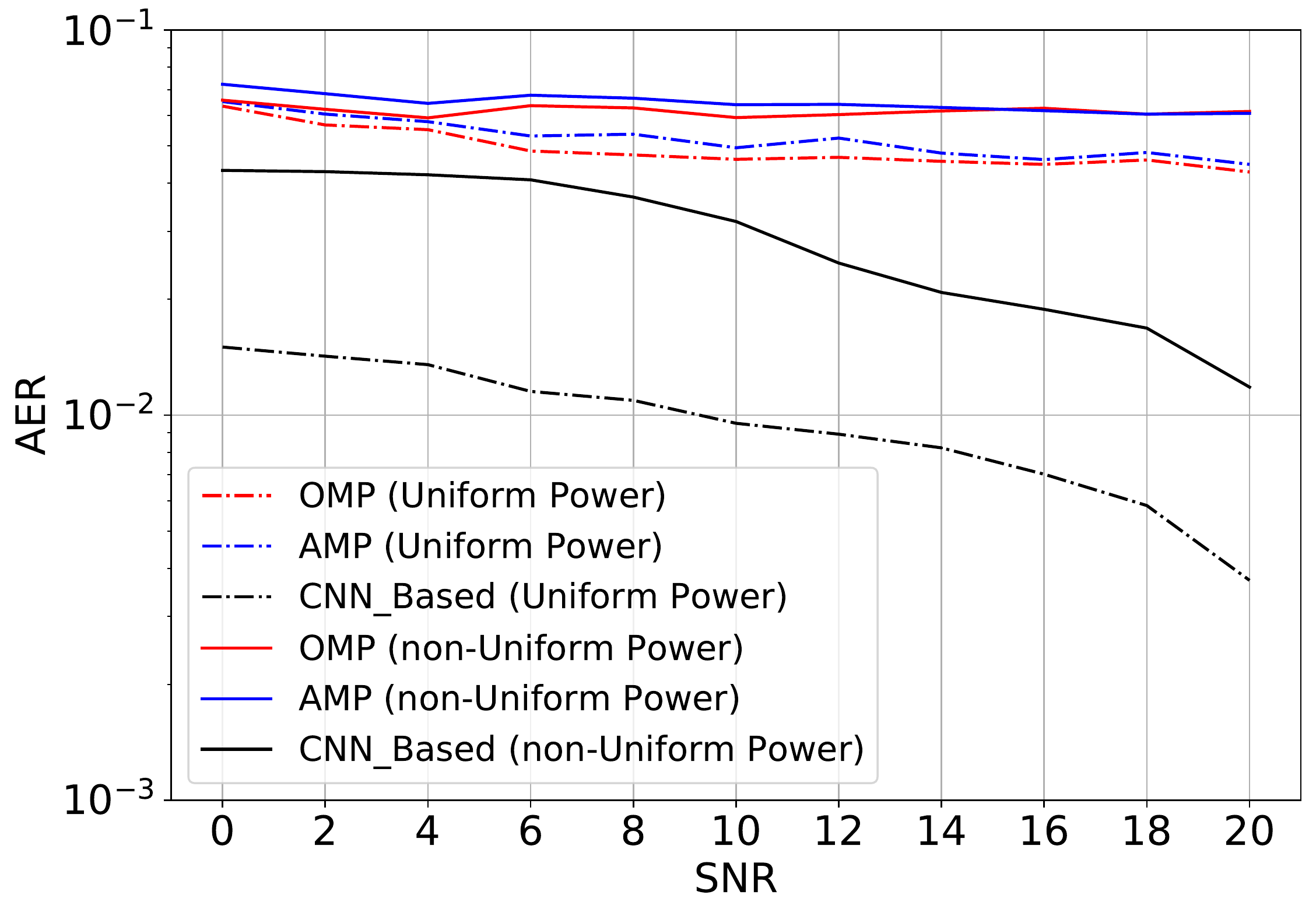}
	\vspace{-.2cm}
	\caption{Achieved BER with MMSE with a priory AD of OMP, AMP, and CNN-AD without knowing the number of active devices.} \label{fig:BER1}	
	\vspace{-.4cm}
\end{figure}
	
\begin{figure}[t]
	\centering
	\vspace{.1cm}
	\includegraphics[width=0.45\textwidth]{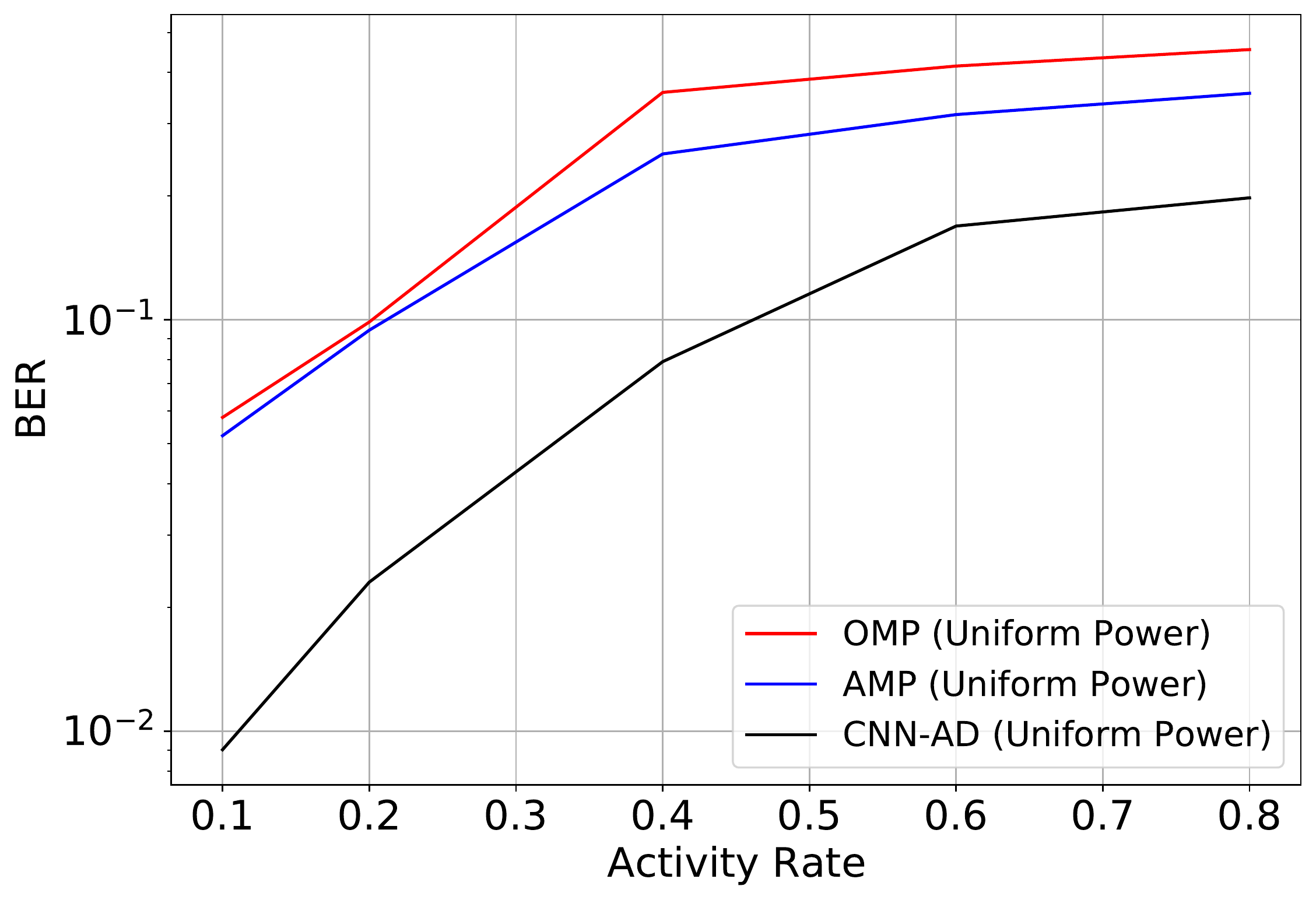}
	\vspace{-.2cm}
	\caption{Impact of $P_{\rm a}$ on the performance of different methods as the priory AD for MMSE in terms of achieved BER.}
	\label{fig:3}	
	\vspace{-.25cm}
\end{figure}
	
\subsection{Simulation Results}
\subsubsection{Performance Evaluation of CNN-AD}
We assess CNN-AD through various simulations and compare it with the existing \ac{cs}-based methods including \ac{omp} \cite{OMP} and \ac{amp}~\cite{AMP}.

The impact of \ac{snr} on the \ac{aer} achieved by different \ac{ad} algorithms in both homogeneous and heterogeneous \ac{iot} networks with uniform and non-uniform power allocation is shown in Fig.~\ref{fig:BER1}.
The \ac{aer} of different methods are compared for a wide range of \ac{snr}s in an \ac{iot} system with total $K=40$ \ac{iot} devices and a single \ac{bs} with $M=100$ receive antennas.
As expected, the \ac{aer} of all \ac{ad} algorithms decreases with increasing \ac{snr}.
However, CNN-AD achieves the best performance since unlike the non-Bayesian greedy algorithms \ac{omp} and \ac{amp}, our method relies on the statistical distributions of device activities and channels and exploit them in the training process.

Fig.~\ref{fig:3} illustrates the effect of activity rate on the \ac{ber} 
for \ac{mmse}-\ac{mud} with different \ac{ad} algorithms at $\gamma=10$ dB \ac{snr}.
As seen, as the activity rate increases, the number of active devices also increases accordingly and thus it becomes difficult to detect all the active devices. This results in a higher \ac{ber}. 
We use $P_{\rm max}=0.1$ to train CNN-AD. Thus, the \ac{mmse}-\ac{mud} with \ac{cnn}-\ac{ad} shows performance degradation for the activity rates of larger than $P_{\rm max}=0.1$. However, it still outperforms the performance of the \ac{mmse}-\ac{mud} with \ac{omp} and \ac{amp} \ac{ad} algorithms.  
It should be mentioned that this performance improves when CNN-AD is trained for a larger value of $P_{\rm max}$.

We further investigate the \ac{ad} algorithms in terms of other metrics for two typical \ac{iot} devices for $P_{\rm max}=0.1$ at $\gamma=10$ dB \ac{snr}, presented in Table \ref{tab:metrics}. 
In this table we compare the precision, recall, and F1-score, defined in \cite{goutte2005probabilistic}, achieved by CNN-AD with \ac{omp} and \ac{amp} \ac{ad} algorithms.
As seen, all metrics are improved by using CNN-AD.

\begin{table}[]
    \centering
    \vspace{.2cm}
    \begin{tabular}{*{2}c | *{3}c}
    \hline
        IoT Device & Model & Precision &  Recall & F1-score\\ \hline
         & OMP & 28\% & 32\% & 30\%\\
         Device A & AMP & 31\% & 35\% & 33\%\\
         & \textbf{CNN-AD} & \textbf{73\%} & \textbf{92\%} & \textbf{81\%} \\ \hline
         & OMP & 33\% & 32\% & 32\%\\
         Device B & AMP & 38\% & 35\% & 36\%\\
         & \textbf{CNN-AD} & \textbf{100\%} & \textbf{83\%} & \textbf{91\%}\\ \hline
    \end{tabular}
    \caption{Performance analysis different algorithms for two typical IoT devices for $P_{\rm max}=0.1$ at $\gamma=10$ dB.}
    \vspace{-.6cm}
    \label{tab:metrics}
\end{table}

    \vspace{-.15cm}
	\section{Conclusions}\vspace{-.15cm}
	\label{Sec:Conclusions}
	In this paper, we consider the problem of \ac{ad} in \ac{iot} networks in grant-free \ac{noma} systems.
	Based on the application, \ac{iot} devices can be inactive for a long period of time and only active in the time of transmission to the \ac{bs}.
	Hence, identifying the active devices is required for an accurate data detection.
	Some studies propose \ac{cs}-based method for \ac{ad}.
	However, high level of message sparsity is necessary for those methods.
	In order to remove this need and exploit the statistical properties of the channels we propose a \ac{cnn}-based method called CNN-AD to detect active \ac{iot} devices.
	Comparison with available methods shows the strength of our algorithm.
	
%

\section*{Acknowledgment}
The study presented in this paper is supported by Alberta Innovates and Natural Sciences and Engineering Research Council of Canada (NSERC).

\bibliographystyle{IEEEtran}
\bibliography{IEEEabrv,Ref}


\end{document}